\documentclass[a4paper,11pt]{article}
\pdfoutput=1 

\usepackage{jheppub} 
\usepackage[T1]{fontenc} 
\usepackage{physics}
\usepackage{float}

\newcommand{\normord}[1]{:\mathrel{#1}:}
\title{Tidal excitation as mixing in thermal CFT}

 \author{Julius Engelsöy}
 \author{and Bo Sundborg}
 \affiliation{The Oskar Klein Centre for Cosmoparticle Physics \& Department of Physics,\\
Stockholm University,\\
AlbaNova, 106 91 Stockholm, Sweden}

\abstract{We use mixed correlators in thermal CFT as clean probes of the strong gravity effects in their holographic duals. 
The dual interpretation of mixing is an inelastic conversion of one field to another field, induced by gravity: tidal excitation. We find an enhanced mixing at high temperatures, corresponding to large AdS black holes, concentrated to small boundary momenta, dual to the deep bulk, where strong gravitational fields are expected. We also find large $\mathcal{O}(1/G_{N})$ tidal conversion in the low temperature phase of the $U(N)$ vector model, strengthening suspicions that the bulk dual of this phase also houses extremely compact objects.}

\begin{document} 
\maketitle
\flushbottom

\section{Introduction}

The possibility to glean otherwise inaccessible strong gravity effects from AdS/CFT has recently received renewed attention. One theme is to apply thermal field theory to the problem. Another is to look for field theory probes that uniquely focus on the bulk regions of interest, be it the singularity \cite{Grinberg:2020fdja,Rodriguez-Gomez:2021pfh}, the region just outside the photon sphere \cite{Dodelson:2020lala}, or the region just inside, which support evanescent modes \cite{ReyRosenhaus2014,JevickiSuzuki2016,JevickiYoon2016,AmadoSundborgThorlaciusWintergerst2017,AmadoSundborgThorlaciusWintergerst2018}. The filtering of information to focus on the respective phenomenon is achieved by, a) studying particular higher dimension operators that only couple in regions of high curvature, b) noting corrections to the propagation of particles from the higher curvature regions, c) finding depletion of elastic scattering in black hole geometries by tidal excitations of the scattered particles, or d) extraction of zero frequency spectral densities on the boundary, corresponding to the propensity of an operator to excite the thermal background (the black hole) by a given momentum. In the last case, the possibility for the operator to deform the background with non-zero momenta while preserving energy is tied to special evanescent modes inside the ISCO (innermost stable circular orbit) for a classical black hole geometry. These modes couple to the boundary only on boundary distances of the order of the inverse temperature, and are drastically suppressed on shorter scales. 

Most of these diverse strong gravity phenomena are related to the phenomenon of mixing in the boundary theory. Is this true for all cases? The high dimension operator in \cite{Grinberg:2020fdja} mixes with multistress tensors, which are directly associated with strong gravity. The two-point correlator in \cite{Rodriguez-Gomez:2021pfh} detects the presence of strong gravity by its modification of the propagation of massive bulk scalars by strong gravity, again represented by multistress tensors in the boundary. The tidal excitation mechanism crucial for removing a bulk-cone singularity (see \cite{Hubeny:2006yu}) in \cite{Dodelson:2020lala} explicitly diffuses the original signal into other signal modes. This mixing of bulk fields is induced by the presence of strong gravity. 
Indeed, tidal stresses have been found to be significant also in horizonless microstate geometries  \cite{TyukovWalkerWarner2018,Bena:2018mpb,Bena:2020iyw,Martinec:2020cml}  that, seen from a distance, resemble black holes. These geometries are also associated with scrambling \cite{Craps:2020ahu}, another probe of strong gravity.
Finally, evanescent modes are related to the strong strong gravity inside the innermost stable orbits, but the relation to mixing is less clear. In this note, we study mixing of a light scalar with the universally present energy density, which is a scalar of the spatial rotation group preserved at finite temperature. If such a mixing coincides with the presence of evanescent modes it would further support the hypothesis that substantial mixing in the boundary theory encodes strong gravity in the bulk.

Aside from the application to strong gravity, thermal CFT is important in its own right. By the introduction of a scale through the inverse temperature $\beta$ the structure of CFTs at finite temperature becomes much richer than at zero temperature. New non-vanishing quantities that characterise finite temperatur physics are thermal one-point functions 
$\left\langle O_i(x) \right\rangle$ and the quantities we focus on, off-diagonal two-point correlators $\left\langle O_i(x) O_j(y) \right\rangle$. There has been substantial interest in thermal conformal correlators recently, see \cite{Iliesiu:2018fao,GobeilMaloneyNgWu2019,Karlsson:2021duj} in addition to the holographically focused papers above. We have not found a general discussion of off-diagonal correlators, however.

Now, let us describe what the observable consequences of mixing of bulk fields in strong gravity would be. First, to simplify the discussion, assume no other mixing mechanism is active. Without mixing a particle traveling through empty spacetime maintains its identity. This statement is open to criticism because the particle concept is elusive in gravitational fields. To be more precise, we consider particle detection and emission from asymptotic regions of spacetime without particle creation, for example in asymptotically AdS or Minkowski geometries, in what is roughly a scattering setup. Without mixing, a particle originating in an asymptotic region remains the same when it returns to an asymptotic region. If it instead passes through a region of strong gravity which tidally excites it, it will appear converted. Of course, the conversion probability will vary with how far the particle has travelled and the effective strength of the gravitational field.

We test the idea to use pure boundary calculations of mixed correlators in simple toy model calculations in large $N$ free vector models, augmented by singlet constraints, and find that such simple models display a rich behaviour. Even though such field theories are simple, they contain rich physics; there is a finite temperature phase transition which in the case of adjoint matter occurs at the AdS scale implying its correspondence with a Hawking--Page transition \cite{Sundborg2000,AharonyMarsanoMinwallaPapadodimasVan-Raamsdonk2004}. For fundamental matter the transition is smoother and is completed at the Planck scale \cite{ShenkerYin2011}. Nevertheless, there is strong evidence---in the form of evanescent modes and the behaviour of 2-point functions---that the transition implies the existence of black hole-like physics in the bulk \cite{AmadoSundborgThorlaciusWintergerst2017,AmadoSundborgThorlaciusWintergerst2018}.

In effect, our model calculation highlights a general property of AdS/CFT with a flavour of string theory\footnote{In string theory, the particle spectrum is sensitive to the background geometry.}, even if the model is not a string theory. These models have instead been argued to be boundary representations  \cite{KlebanovPolyakov2002} of higher spin gravity \cite{FradkinVasiliev1987b,FradkinVasiliev1987a}, and constitute classically closed sub-sectors \cite{SezginSundell2002} of tensionless string theory \cite{Sundborg2001}. Mixed correlator calculations similar to those  we present in this work can be repeated in matrix models which are directly related to string theory. In spite of the technical similarities, we expect the results to differ between models, as observed for other correlators in \cite{AmadoSundborgThorlaciusWintergerst2017,AmadoSundborgThorlaciusWintergerst2018}. We believe that the compositeness of the singlet boundary operators is crucial for the similarities in the gravitational interpretation. It implies that the bulk fields represent different states of the same object. This is clearly the case, by a different mechanism, in worldsheet formulations of string theory. The widely different state counting and the more complicated composition of matrix representations cause differences.

In section \ref{sec:op} we discuss the subtleties in defining thermal composite operators in large $N$ singlet models and their concrete representation. Section \ref{sec:corr} explains the central calculations of mixed correlators and their thermal behaviour, while section \ref{sec:disc} discusses the physics of our results and our general conclusions. 

\section{Composite operators in thermal equilibrium}\label{sec:op}
To study thermal correlation functions, we need to specify for what operators we should compute correlators. The answer is deceptively simple. At least in the absence of phase transitions, we should consider an operator $O_{\beta}$ at temperature $\beta^{-1}$ for each operator $O$ in vacuum. Often,  we do not even need to make a distinction between $O_{\beta} $ and $O$ and there is little to discuss. This would be the case in bootstrap approaches to thermal CFT. The formalism will take care of the rest. In our framework we wish to describe the physical operators in terms of hidden large $N$ constituents, which provide the building blocks for calculating correlation functions. The constituents are hidden by requiring that physical states and operators are singlets. This is achieved by introducing a formal gauge symmetry and a corresponding  formal vector potential to represent the projection onto the singlet sector of the original theory. Technically, a non-trivial zero mode of a gauge potential is physical because the thermal circle $S_{\beta}$ is non-contractible. That said, the zero mode is still non-dynamical, and its path integral produces the singlet condition.
The singlet condition at non-zero temperatures will turn out to be more interesting for heavier operators (with more gauge covariant derivatives in their definition) than for the lightest states. The intuitive idea to keep in mind is that a composite object at thermal equilibrium is also in an internal equilibrium at the same temperature, and should be tuned accordingly. There is more adaption for heavier operators with more internal structure.

\subsection{The model}
Before uncovering the subtler issues above we should specify our model. We consider a $U(N)$ vector model with scalars in the fundamental (and the anti-fundamental) representation. Denoting them by $\varphi$ and $\varphi^{\dagger}$
the lowest dimension singlet scalar in the theory,
$$
O = \varphi^{\dagger}\varphi
$$
is the ground state operator. It is important that $O=J^0$ is the lightest operator in \emph{a tower of related operators} 
$$
J^s_{\mu_1...\mu_s}= \sum_{k=0}^{s} a_{s k}\ \partial_{\{\mu_1}...\partial_{\mu_k } \varphi^{\dagger}{\partial}_{\mu_{k+1}}...\partial_{\mu_s\}} \varphi , 
$$
with the curly brackets symmetrising and projecting out traces. With appropriate choices of coefficients $a_{s k}$
these operators are conserved currents,
$$
\partial^\mu J^s_{\mu \mu_1...\mu_s} =0
$$
due to their construction in terms of free fields. The free field constituents provide the basis for the higher spin symmetry generated by the currents. The conserved currents are dual to bulk massless higher spin gauge fields. For example, the energy-momentum tensor $J^2_{\mu\nu}=T_{\mu\nu}$ is dual to the bulk graviton, as usual in AdS/CFT.

We have written down expressions for this free vector model in flat space, but with appropriate generalisations we can map flat spacetime to a cylindrical spacetime $S^{d-1}\times\mathbb{R}$. A standard way to obtain thermal equilibrium correlators is then to apply the imaginary time formalism, which considers the Euclidean version of the field theory with a circular time dimension of circumference $\beta$ equal to the inverse temperature. In our case, the Euclidean space is $S^{d-1}\times S_{\beta}$ and the ratio of the radii of the spatial sphere and the time circle is conformal invariant. This invariant is thus directly related to temperature.
\subsection{The thermal operators}

We thus define composite operators as certain (homogeneous) polynomials of operators, and compute their correlators by Wick contractions. In the case of vector models the operators are the quadratic expressions above, but the idea is the same for matrix models which may represent tensionless limits of string theory \cite{Haggi-ManiSundborg2000}. We  impose singlet conditions on physical operators, just as for physical states. This machinery secretly involves ordering prescriptions for the polynomial operators, as well as mechanisms to ascertain the residues of gauge invariance. Technically, this means including the formal gauge zero mode in the construction of gauge invariant thermal higher spin currents. We may recall lattice definitions of charged fields and covariantly transforming currents as examples of complications analogous to those arising for singlet composite operators due to combined requirements on ordering, locality  and invariance. In our case, an integral over the gauge zero mode will eventually perform the projection onto the singlet sector. Since this projection integral is affected by temperature, temperature dependence is introduced in the singlet operators themselves. Once we have constructed the operators $J_{\beta}^s$ discussed above, with these subtleties in mind, we can remove the $\beta$ subscript, which was just a reminder of the essential internal adjustments in composite objects. As will be seen below, these operators are formally gauge singlet operators constructed with gauge covariant derivatives, but as emphasised above, the gauge fields in the formal expressions just prepares the operators for the requisite projection integrals that tunes their internal structure to thermal equilibrium.\footnote{For some models, like the maximally supersymmetric Yang-Mills matrix model, the formal gauge potential can be viewed as a part of the dynamical vector potential which completes the model to thermal super-Yang-Mills theory, whether the gauge coupling vanishes or is non-zero.}

In this first study of mixing in singlet models/models with higher spin symmetry, we confine our attention to two of the simplest operators that can mix, the ground state scalar and the energy density, which is the time-time component of the stress-energy tensor, dual to bulk gravitons and undoubtedly important in gauge-gravity duality.

We need an expression for the stress--energy tensor. For complex scalars coupled to a gauge field and to the curvature scalar $R_{\xi}$ we have the conformally improved stress-energy tensor
\begin{align}\label{eq:stress}
T_{\mu\nu} &= D_{\mu}\varphi^{\dagger}D_{\nu}\varphi+D_{\nu}\varphi^{\dagger}D_{\mu}\varphi-g_{\mu\nu}D_{\alpha}\varphi^{\dagger}D^{\alpha}\varphi + 2\xi(G_{\mu\nu}+g_{\mu\nu}\Box-D_{\mu}D_{\nu})\abs{\varphi}^{2}
\end{align}
where $\xi = (d-2)/(4(d-1))$, $G_{\mu\nu}$ is the Einstein tensor, all geometric quantities are to be computed on $S^{d-1}\times S_{\beta}$ and we used $D_\mu$ to denote the gauge \emph{and} generally covariant derivative. 
With this notation we can use Leibniz rule freely, and the equations of motion both read $(D^{2}-\xi R_{\xi})\varphi = 0= (D^{2}-\xi R_{\xi})\varphi^{\dagger}$ even if the covariant derivative contain different terms for the fundamental and anti-fundamental representations. Applying these relations to minimise the presence of second derivatives from the d'Alembertian, we find
\begin{align}
T_{\mu\nu} 
&=D_{\mu}\varphi^{\dagger}D_{\nu}\varphi+D_{\nu}\varphi^{\dagger}D_{\mu}\varphi-g_{\mu\nu}D_{\alpha}\varphi^{\dagger}D^{\alpha}\varphi + 2\xi(G_{\mu\nu}-D_{\mu}D_{\nu})(\varphi^{\dagger }
\varphi)\notag\\
&+ 2\xi g_{\mu\nu}\left(2D_{\alpha}\varphi^{\dagger}D^{\alpha}\varphi+2\xi R_{\xi}\varphi^{\dagger}\varphi \right).
\end{align}
The 00-component reads
\begin{align}
T_{00}&= (1+4\xi)\partial_{0}\varphi^{\dagger}\partial_{0}\varphi -(1-4\xi)g_{ij}\partial_{i}\varphi^{\dagger}\partial_{j}\varphi - 2\xi((\partial_{0}^{2}\varphi^{\dagger})\varphi+ 2\partial_{0}\varphi^{\dagger}\partial_{0}\varphi+\varphi^{\dagger}\partial_{0}^{2}\varphi)\notag\\
&+ 2\xi (G_{00}+2\xi R_{\xi})\abs{\varphi}^{2} +i(1+4\xi) (\partial_{0}\varphi^{\dagger}\alpha\varphi - \varphi^{\dagger}\alpha\partial_{0}\varphi)+ (1-4\xi) \varphi^{\dagger}\alpha^{2}\varphi,
\end{align}
where $\alpha$ is the gauge potential zero mode in the covariant derivative, $D_0 \varphi = (\partial_0 + i \alpha)\varphi$.
For $d=4$ we have $\xi = 1/6$, $G_{00} = 3/R^{2}$, $R_{\xi} = 6/R^{2}$ such that
\begin{align}  \label{eq:correctedT00}
T_{00} &=  \partial_{0}\varphi^{\dagger}\partial_{0}\varphi - \frac{1}{3}(g_{ij}\partial_{i}\varphi^{\dagger}\partial_{j}\varphi+(\partial_{0}^{2}\varphi^{\dagger})\varphi+\varphi^{\dagger}\partial_{0}^{2}\varphi)+ \frac{5}{3R^{2}}\abs{\varphi}^{2} \notag\\
&+\frac{5i}{3} (\partial_{0}\varphi^{\dagger}\alpha\varphi - \varphi^{\dagger}\alpha\partial_{0}\varphi)+ \frac{1}{3} \varphi^{\dagger}\alpha^{2}\varphi
\notag\\
&\equiv T^{\mathrm{(st)}}_{00} +\frac{5i}{3} (\partial_{0}\varphi^{\dagger}\alpha\varphi - \varphi^{\dagger}\alpha\partial_{0}\varphi)+ \frac{1}{3} \varphi^{\dagger}\alpha^{2}\varphi \equiv T^{\mathrm{(st)}}_{00} +T^{\mathrm{(th)}}_{00} \,,
\end{align} 
where we have defined $T^{\mathrm{(st)}}_{00}$ as the ``standard" energy density expression, and $T^{\mathrm{(th)}}_{00}$ as the thermal improvement term, which when averaged over gauge potential zero mode readjusts the equilibrium composite energy density operator appropriately.

\section{Thermal CFT correlators}\label{sec:corr}

At finite temperature, some basic properties of conformal field theory correlators change. The simplest change is that one-point functions, previously required to vanish by conformal invariance, now can acquire non-zero values. By the still remaining translation and time-translation invariances, the one-point functions are constants, only depending on which operator is studied, and on temperature. They may thus contain slightly more information than the free energy, since there is one one-point function per operator. Two-point functions contain much more information, as the example of diagonal two-point functions demonstrate: spacetime dependence is added to temperature dependence, although still heavily constrained by translation invariances. In contrast, at zero temperature normalisation, spin and conformal dimension completely fix diagonal two-point functions. Off-diagonal two-point functions are thus particularly interesting as the simplest spacetime dependent quantities which specifically characterise non-zero temperature. After some more background we begin with the study of extremely high temperatures, consider the general case, and then specialise further to the lowest temperatures and to more typical temperatures in the low temperature phase.

\subsection{Background}

The singlet constraint can be enforced by a gauge field whose coupling is taken to zero in a limiting procedure. The gauge sector and the matter sector then decouple apart from the appearance of a temporal gauge zero mode in the matter partition function. 

For concreteness we consider thermal mixing in the $U(N)$ vector model, but the calculation can be readily extended to other gauge groups and representations. The matter sector Euclidean action is then
\begin{equation}\label{eq:action}
S = -\int\dd^{d}x\, \varphi^{\dagger}\left((\partial_{\tau}+i\alpha)^{2}+\partial_{i}^{2} - \frac{(d-2)^{2}}{4R^{2}}\right)\varphi\,,
\end{equation}
where $\alpha$ is the temporal gauge zero mode and $R$ is the radius of the spatial sphere the theory lives on. We will use units in which $R=1$, and thus seldom write it out. We have also suppressed the $U(N)$ indices which will be denoted by lowercase Latin letters when needed.

We now compute the mixed 2-point function between the quadratic scalar composite operator and the energy density In the method we employ we will first need the scalar--scalar 2-point function. It can be obtained by Wick contracting fundamental propagators $\Pi_{ab}(\tau,\theta)$ obtained from \eqref{eq:action}. The derivation of the fundamental propagator is sketched in appendix \ref{app:fundamental}.

We find the scalar--scalar 2-point function:
\begin{align}
\expval{\abs{\varphi}^{2}(\tau,\theta)\abs{\varphi}^{2}(0,0)} &= \Pi_{ab}(\tau,\theta)(\Pi_{ab}(\tau,\theta))^{*} \notag\\
&= \sum_{i=1}^{N}\sum_{n,m=-\infty}^{\infty}\frac{e^{- i (n-m) \lambda_{i}}}{(\cosh(\tau+\beta n)-\cos\theta)^{\sigma}(\cosh(\tau+\beta m)-\cos\theta)^{\sigma}}\,,
\end{align}
defining $\sigma \equiv (d-2)/2$. Here, the $\lambda_i$ are rescaled eigenvalues of the temporal vector potential $\alpha$ taking values between $-\pi$ and $\pi$. For $N\gg 1$ we may approximate the eigenvalues as being part of a continuous distribution $\rho(\lambda)$ and hence, rewrite this expression as
\begin{align}\label{eq:sc2}
\expval{\abs{\varphi}^{2}(\tau,\theta)\abs{\varphi}^{2}(0,0)} &= \int_{-\pi}^{\pi}\dd \lambda \,\rho(\lambda)\sum_{n,m=-\infty}^{\infty}\frac{e^{- i (n-m) \lambda}}{(\cosh(\tau+\beta n)-\cos\theta)^{\sigma}(\cosh(\tau+\beta m)-\cos\theta)^{\sigma}}\,.
\end{align}
The large $N$ limit also allows us to employ a saddle point approximation of $\rho(\lambda)$. In short, the eigenvalue distribution takes a near constant form in the low temperature phase and approaches a delta function far above the phase transition \cite{Sundborg2000,AharonyMarsanoMinwallaPapadodimasVan-Raamsdonk2004,ShenkerYin2011,AmadoSundborgThorlaciusWintergerst2017,AmadoSundborgThorlaciusWintergerst2018}. For the particular case of the vector models \cite{ShenkerYin2011} the critical temperature interestingly scales with $N$ as $T_c \sim N^{1/(d-1)}$, which is the Planck temperature in the dual higher spin gravity theories. It is important to note the increasing range of temperatures between the AdS temperature $T_{\mathrm{AdS}}=T_R = 1$ and $T_c$ in the large $N$ limit.

\subsection{Mixing at extremely high temperatures}

In the high extreme high temperature limit the expressions look more familiar, which is why we explain this calculation first. By extremely high temperatures, we mean the regime in which may use a delta function eigenvalue density. In other words, we use that \emph{the gauge zero mode vanishes at the saddle point.} We may then obtain the sought-after 2-point function $\expval{T_{00}(\tau,\theta)\abs{\varphi}^{2}(0,0)}$ by constructing a point-splitting differential operator \cite{Giombi:2016ejx} and letting it act on \eqref{eq:sc2}. This operator, in turn, can be obtained from the expression of the conformally improved stress--energy tensor \eqref{eq:stress} now specialised to the case without gauge potential: 
\begin{equation}
T_{\mu\nu} = \partial_{\mu}\varphi^{\dagger}\partial_{\nu}\varphi+\partial_{\nu}\varphi^{\dagger}\partial_{\mu}\varphi-g_{\mu\nu}\partial_{\alpha}\varphi^{\dagger}\partial^{\alpha}\varphi + 2\xi(G_{\mu\nu}+g_{\mu\nu}\Box-\nabla_{\mu}\nabla_{\nu})\abs{\varphi}^{2}\,,
\end{equation}
In this limit $T_{\mu\nu}=T^{\mathrm{(st)}}_{\mu\nu}$.

For better analytic control we choose to work out the toy model in $d=4$ and in the short distance limit (or equivalently in the large sphere limit, if we allow $R$ to vary). There, the energy density becomes
\begin{align}\label{eq:Tflatsph}
T_{00} &\simeq \partial_{0}\varphi^{\dagger}\partial_{0}\varphi - \frac{1}{3}\Bigg(\partial_{x}\varphi^{\dagger}\partial_{x}\varphi  + \partial_{y}\varphi^{\dagger}\partial_{y}\varphi + \partial_{z}\varphi^{\dagger}\partial_{z}\varphi + (\partial_{0} \partial_{0}\varphi^{\dagger} )\varphi + \varphi^{\dagger} (\partial_{0} \partial_{0}\varphi)\Bigg)\notag\\
&\equiv \lim_{u\rightarrow v} D(u,v)\varphi^{\dagger}(u)\varphi(v)\,, 
\end{align}
where we have defined the the point-splitting differential operator $\lim_{u\rightarrow v} D(u,v)$ and $u,v$ refer to the respective arguments of the fundamental field operators.
Furthermore, \eqref{eq:sc2} reduces to
\begin{align}\label{eq:highT}
\expval{\abs{\varphi}^{2}(\tau,x)\abs{\varphi}^{2}(0,0)} &\simeq \frac{1}{4}\left(\sum_{m=-\infty}^{\infty}\frac{1}{(\tau+\beta m)^{2}+x^{2}}\right)^{2}\notag\\
&= \frac{\pi^{2}}{2x^{2}\beta^{2}}\frac{\sinh^{2} \tfrac{2\pi x}{\beta}}{(\cos \tfrac{2\pi \tau}{\beta}-\cosh \tfrac{2\pi x}{\beta})^{2}}\equiv G(\tau,x)\,,
\end{align}
where we have substituted for the high temperature saddle point eigenvalue distribution $\rho(\lambda) = \delta(\lambda)$.  

Now applying the differential operator defined in \eqref{eq:Tflatsph} and keeping track which arguments belong to which fundamental field in order to correctly perform the point splitting yields a rather complicated expression for the mixed 2-point function, but its high temperature limit is simple,
\begin{align} \label{eq:extreme_high}
   \expval{T_{00}(\tau,x)\abs{\varphi}^{2}(0,0)} \to -\frac{2 \pi ^2}{3 \beta ^2 x^4}.
\end{align} 
We may Fourier transform the full expression to investigate the UV and IR behavior. To investigate static configurations and for further analytic control we compute the zero mode
\begin{align}\label{eq:zero}
&G(\omega_{n}=0,x) = \int_{0}^{\beta}\dd\tau\,G(x,0) \notag\\
&= -\frac{\pi^{2}}{12x^{4}\beta^{3}\sinh^{2}\tfrac{2\pi x}{\beta}}\left[\beta\left(8\pi x + \beta\frac{\cosh\tfrac{6\pi x}{\beta}}{\sinh\tfrac{2\pi x}{\beta}}\right)-(32\pi^{2}x^{2}+\beta^{2})\coth\tfrac{2\pi x}{\beta}\right]\,,
\end{align}
This expression can be divided into the two asymptotic regions
\begin{equation}\label{eq:appzero}
G(\omega_{n}=0,x) \approx \begin{cases}-\frac{64\pi^{5}}{135\beta^{4}x} & x\lesssim\frac{3^{2/3}5^{1/3}\beta}{4\pi}\,,\\
-\frac{\pi^{2}}{3x^{4}\beta}& x\gtrsim \frac{3^{2/3}5^{1/3}\beta}{4\pi}\,.\end{cases}
\end{equation}
We note that this correlator vanishes at zero temperature equivalent to $\beta\rightarrow\infty$ as expected. We may now proceed to Fourier transform the $x$ coordinate according to
 \begin{align}\label{eq:xint}
G(\omega_{n}=0,k) &= 2\pi \int_{0}^{\infty}\dd x\,x^{2} \int_{0}^{\pi}\dd\theta\, e^{-i k x \cos \theta}G(\omega_{n}=0,x)\notag\\
&=4\pi \int_{0}^{\infty}\dd x \,x \,\frac{\sin k x}{k}\,G(\omega_{n}=0,x)\,,
\end{align}
where we use the fact that $G_{T_{00}\varphi^{2}}(\omega_{n}=0,x)$ is independent of the azimuthal angle. We have not been able to analytically compute this integral of \eqref{eq:zero} but one can perform the integral numerically and compare to the approximation using \eqref{eq:appzero}. The sharp crossover scale of the approximation produces oscillations in the Fourier transform but their envelope agrees with the smoother numerical result presented below. The result of \eqref{eq:xint} has the form
\begin{equation}\label{eq:fdef}
G(\omega_{n}=0,k) = -\frac{1}{\beta^{2}}f(k\beta)\,,
\end{equation}
where $f(x)$ is a positive definite dimensionless function of the dimensionless combination $k\beta$. A log-log plot of the numerical result for $f(k\beta)$ can be seen in Fig. \ref{fig:plot} below. There is a crossover scale of the order $k \sim 1/\beta$ indicating that as we move down from the UV into the IR, or correspondingly inwards in the dual gravitational bulk, the strength of the mixing increases as a power-law until the energy scale $k$ reaches the crossover value set by the temperature of the putative black hole. Below the crossover, it approaches a constant value of maximum mixing.

\begin{figure}[H]
  \centering
    \includegraphics[scale=0.7]{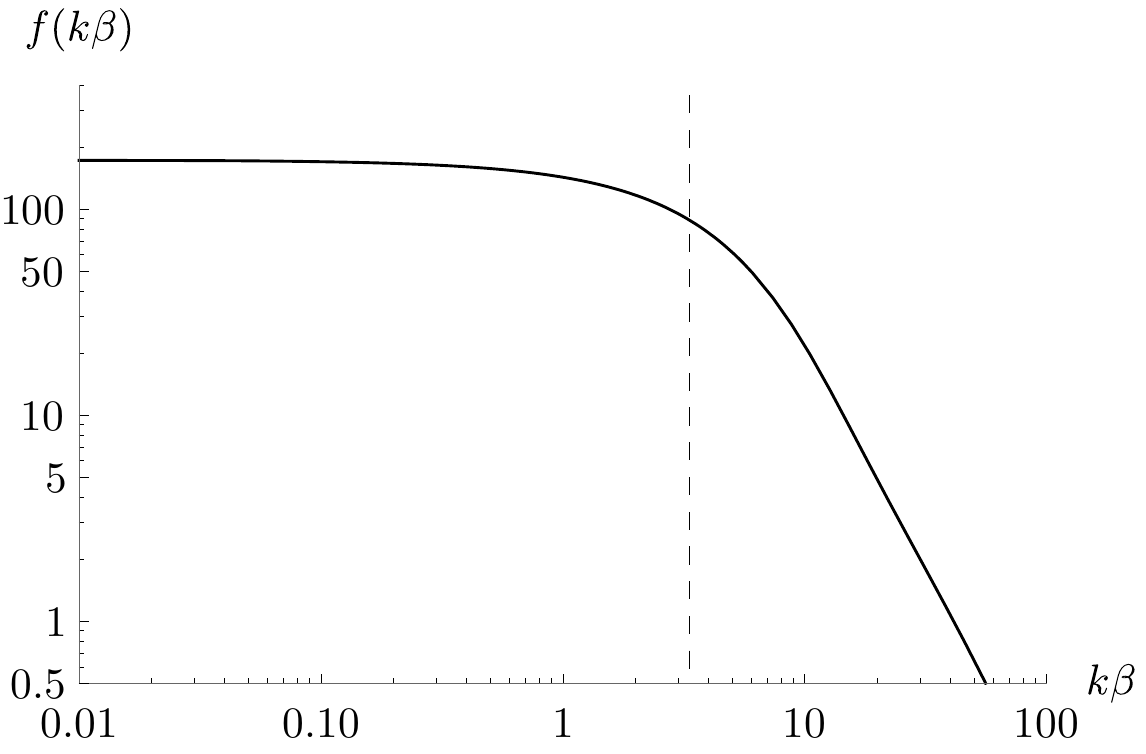}
      \caption{Log-log plot of the momentum space static mixed correlator $f(k\beta)$ defined in \eqref{eq:fdef}. The dashed line indicates the crossover scale. The plateau means large conversion in the central region of the bulk, and the power law fall off means that conversion is suppressed outside this central hotspot.}
      \label{fig:plot}
\end{figure}

\subsection{The general mixing correlator}
We see from the expression for the energy density \eqref{eq:correctedT00} that the mixed correlator for general temperatures,
$$
\expval{T_{00}(\tau,x)\abs{\varphi}^{2}(0,0)},
$$
receives a correction depending on $\alpha$, the vector potential zero mode. The detailed derivation of the correction term $\expval{T_{00}^{(th)}(\tau,x)\abs{\varphi}^{2}(0,0)}$ is deferred to appendix \ref{app:correction}. 

The standard term  $\expval{T_{00}^{(st)}(\tau,x)\abs{\varphi}^{2}(0,0)}$ is obtained, as before, from the application of the point-splitting differential operator in eq. \eqref{eq:Tflatsph} on the scalar-scalar propagator \eqref{eq:sc2}, which actually annihilates this contribution completely for constant eigenvalue density and first order in $\theta$ and $\beta \ll 1$. These conditions are compatible with the low temperature phase, provided the critical point is not approached. 

For the same typical temperature in the low temperature phase, where the eigenvalue density is approximately constant, we can rewrite eq. \eqref{eq:th_uniform_density} for the thermal correction 
\begin{align}\label{eq:low-th}
\expval{T^{(th)}_{00}(\tau,x) O(0,0)} &\simeq \frac{11}{3\beta^{2}} \sum_{n,m=-\infty}^{\infty}\frac{I_2(m-n)}{(\cosh(\tau+\beta n)-\cos\theta)^{\sigma}(\cosh(\tau+\beta m)-\cos\theta)^{\sigma}}\notag\\
&+\frac{10}{3\beta} \sum_{n,m=-\infty}^{\infty}\frac{\sigma \sinh(\tau+\beta n)\Im I_1(n-m)}{(\cosh(\tau+\beta n)-\cos\theta)^{\sigma+1}(\cosh(\tau+\beta m)-\cos\theta)^{\sigma}}\,.
\end{align}
with the help of elementary eigenvalue integrals,
\begin{align}
I_1(m)&\equiv \int_{-\pi}^{\pi} \mathrm{d}\lambda \lambda e^{i m\lambda} =-\frac{2 \pi i }{m} (-1)^m,\ I_1(0)=0,
\\
I_2(m)&\equiv \int_{-\pi}^{\pi} \mathrm{d}\lambda \lambda^2 e^{i m\lambda}  \frac{4 \pi}{m^2}(-1)^m,
\\
I_2(0) &=\frac{2\pi^3}{3}.
\end{align}

A more symmetric expression is
\begin{align}
\expval{T^{(th)}_{00}(\tau,x) O(0,0)} &\simeq \frac{11}{3\beta^{2}} \sum_{n,m=-\infty}^{\infty}\frac{I_2(m-n)}{D_{m,n}(\beta;\tau,\theta)^{\sigma}}\notag\\
&+\frac{5}{3\beta} \sum_{n,m=-\infty}^{\infty}\frac{N_{m,n}(\beta;\tau,\theta)\Im I_1(n-m)}{D_{m,n}(\beta;\tau,\theta)^{\sigma}}\label{eq:th_below_mixing}
\\
D_{m,n}(\beta;\tau,\theta)&\equiv \frac{1}{2} (\cosh (\beta (m-n) )+\cosh (\beta (m+n) +2 \tau))\notag\\
&-2 \cos \theta  \cosh \frac{\beta (m-n) }{2} \cosh \frac{\beta (m+n) + 2\tau}{2}+\cos ^2 \theta \label{eq:denominator}
\\
N_{m,n}(\beta;\tau,\theta)&\equiv \frac{\sinh(\tau+\beta n)}{(\cosh(\tau+\beta n)-\cos\theta)} -\frac{\sinh(\tau+\beta m)}{(\cosh(\tau+\beta m)-\cos\theta)}\notag\\
&= \frac{2 \sinh \frac{1}{2} \beta  (n-m) \left(\cosh \frac{1}{2} \beta  (n-m) -\cos \theta  \cosh \frac{1}{2} (\beta  (m+n)+2 \tau )\right)}{D_{m,n}(\beta;\tau,\theta)}\,.
\end{align}
The above form of the correlator is complicated but useful for further processing. We have reinstated general dimensions via $\sigma \equiv (d-2)/2$, since these expressions do not simplify considerably in $d=4$.

\subsection{Cold mixing correlators}
The simplest limit of the above sums appears for temperatures lower than AdS temperatures, $\beta \gtrsim 1$. In this regime, all terms except the $m=n=0$ terms are exponentially damped. In particular, $m=n$ and the second series in eq. \eqref{eq:low-th} does not contribute at all. We get 
\begin{align}\label{eq:cold}
\left\langle T_{00}^{(th)}(\tau,\theta)\, O(0,0) \right\rangle &\simeq \frac{22\pi^3}{9\beta^{2}} \frac{1}{\left((\cosh\tau-\cos\theta)(\cosh\tau-\cos\theta)\right)^{\sigma}}\,.
\end{align}
Note that this term depends on coordinates like a vacuum term, but is multiplied by the thermal factor $\beta^{-2} \lesssim 1$. Indeed, it is functionally the same as the scalar-scalar correlator in this approximation, only suppressed by the small temperature dependent factor.

\subsection{Warm mixing correlators below the critical temperature}

In most of the vector model low temperature phase, many terms with $m \neq 0 \neq n$ contribute, corresponding to the excitation of higher modes of the charged constituent scalars. If $\beta_c \ll \beta \ll 1$ another approximation of the mixing correction \eqref{eq:th_below_mixing} is instead useful. Since in vector models $\beta_c \to 0$ as $N \to \infty$ this is an important ``low temperature phase" regime, even though the temperature is large in our units, fixed by the curvature scale. We note that the denominator \eqref{eq:denominator} controls convergence of the sums, and that convergence is preserved by keeping quadratic terms in the limits $\beta m \to$ 0 and $\beta n \to 0$ with $0 <\tau<\beta$. Hence, we get well defined leading expressions approximating the exact result by  taking the limit of $N_{m,n}$ and expanding $D_{m,n}$ to quadratic order.

Now, rewriting the angular dependence in terms of 
$$
\Theta \equiv \sin{\frac{\theta}{2}}
$$
we find the explicit low temperature phase expressions
\begin{align}
\left\langle T_{00}^{(th)}(\tau,\theta)\, O(0,0) \right\rangle &\simeq \frac{11}{3\beta^{2}}  
\bigg(
\sum_{m=-\infty}^{\infty}
\frac{2\pi^3}
{3 \left[4\Theta^4 +2 \Theta^2 (\beta  m+ \tau )^2\right]^{\sigma}}
\notag\\
&
+\sum_{n \neq m}
\frac{ (-1)^{(m-n)} 4 \pi}
{(m-n)^2\left[4\Theta^4 + \Theta^2\left(\frac{1}{2} (\beta  (m+n)+2 \tau )^2+\frac{1}{2} (\beta  (m-n))^2\right)\right]^{\sigma}}
\bigg)
\notag\\
&
-\frac{5}{3} 
\sum_{n\neq m}\frac{ (-1)^{(m-n)} 4 \pi  \Theta^2 }{ \left[4\Theta^4 + \Theta^2\left(\frac{1}{2} (\beta  (m+n)+2 \tau )^2+\frac{1}{2} (\beta  (m-n))^2\right)\right]^{\sigma+1}}
\,.
\end{align}

The first term above can be evaluated in $d=4$ to yield
\begin{align}\label{eq:aligned}
\left\langle T_{00}^{(th)}(\tau,\theta)\, O(0,0) \right\rangle &=\frac{11 \pi ^3 \arctan \left(-\frac{\sqrt{2} \Theta }{\tau }\right)}{18 \sqrt{2} \beta ^3 \Theta }.
\end{align}
Interestingly, we have found a term in a large intermediate temperature regime growing more rapidly with temperature than the extreme low temperature mixing, or the extreme high temperature mixing.

We have investigated the other terms, dividing them into more sub-terms, and have not yet evaluated them, but instead determined their scaling properties for $\beta \ll 1$. In short, no term grows more rapidly with temperature than the term above, but the circumstance that their angular dependence are different allows us to conclude without further evaluations that the sum of $\beta^{-3}$ terms is non-vanishing.

\section{Discussion and conclusion}\label{sec:disc}

\subsection{The different faces of conversion}
There are different contributions to thermal mixing with different physical origin and different thermal profiles. This is expected, but our detailed investigation of a concrete case provides a template for what can happen in models of fields built of other constituents. The singlet models under study can be viewed as vanishing coupling limits of large $N$ matter charged under the gauge group. Such a limit is intuitive but formal, and a construction of the precise path integral for these models may be better. We have implicitly used this picture here, by insisting that the ultimate physical interpretation is free of the gauge fields that help building the singlet spectrum we are interested in. This may be compared with a bootstrap approach, which is more general, but where the picture of constituents is obscure. In practice, the direct path integral definition involves integration over projection gauge fields with zero field strength along non-contractible loops in spacetime, and vector or matrix charged matter which is free except for the coupling to the projection gauge fields.

The identification of the proper thermal operators is essential for our calculation of the mixing of composite operators. The importance of compositeness is a familiar theme in investigations of fundamental holography \cite{BanerjeePapadodimRajuSamantrayShrivasta2020,Banerjee:2019iwd}, but we encountered a new twist. The operators of the singlet models have to be singlets under the gauge group, and this condition necessitates replacing partial derivatives with gauge covariant derivatives, as usual. The consequences are however quite non-trivial for excited composite operators at finite temperature. Due to the non-contractible curves in the imaginary time description of thermal field theory, completely new terms contribute to the proper definition of the higher dimension thermal composite operators, such as the energy density operator.

We can divide the stress tensor into terms that are the standard operator expressions $T^{\mathrm{(st)}}_{\mu\nu}$ and the thermal correction term $T^{\mathrm{(th)}}_{\mu\nu}$. 
The correlators with the ground state scalar will then be divided similarly. We have found that these two terms have somewhat complementary mixing properties: the mixing of $T^{\mathrm{(st)}}_{\mu\nu}$ with the scalar essentially vanishes below the critical temperature, while the mixing of  $T^{\mathrm{(th)}}_{\mu\nu}$ with the scalar vanishes in the high temperature limit, defined by vanishing holonomy eigenvalues. We have proposed the interpretation that the correction term $T^{\mathrm{(th)}}_{\mu\nu}$ performs the internal readjustments which are necessary for a composite object in thermal equilibrium. Then, the high temperature limit should effectively liberate the constituents. This is indeed the case, as demonstrated by the effective thermodynamic degrees of freedom detectable through the scaling of the free energy with $N$ above the critical temperatures. 

Although both stress tensor terms are essential for the invariance of the thermal stress tensor, 
it is far from obvious to us how the full expression is simpler than its parts. We calculate the mixing of the two terms separately and only when there are similar contributions from the two terms do we exercise special caution, since there could be cancellations. 

\subsection{The temperature evolution of mixing}
There are two temperature scales in vector models on the sphere (and higher spin theory in AdS). The lowest scale, $T=T_R=1$ in our units, is simply the curvature scale on the sphere or in AdS, and the high scale $T=T_c= \alpha N^{1/(d-1)}$ is a scale similar to the Hagedorn scale in superstring theory in AdS \cite{Hagedorn:1965st,Sundborg:1984uk,Sundborg2000,HarmarkWilhelm2018,Harmark:2018red} in that it tracks when states starts to get over-counted by simple Fock space counting. From a dynamical perspective, the phase transition in  vector models is even more interesting since it takes place at the Planck scale set by the gravitational coupling corresponding to Newton's constant in higher spin gravity. 

The lower scale, $T_R$, is not associated with a phase transition, but important nonetheless, and corresponds to the recurrence time for vacuum processes and the complementary quantised frequencies. At lower temperatures, higher dimension operators are suppressed in the heat bath, and propagation of the ground state scalar is virtually unhindered, but the number of effectively available states is small. This cold system is essentially frozen. Probing the cold system by correlators is very similar to probing the vacuum, except with regards to mixing. We found in eq. \eqref{eq:cold} that mixing is non-zero already in this cold environment, and proportional to $T^2$. Since $T < 1$, this temperature factor is still a suppression factor compared to the scalar-scalar correlator, which is of the same functional form. Thus, one interpretation of the cold system is simply that cold suppresses mixing. Conversely, there is actually noticeable mixing in the free vector model already at temperatures of the order of the curvature scale. 

How can we reconcile this noticeable conversion already at relatively low temperatures with the idea that conversion should probe strong gravity effects? The mixing in the boundary correlators tests the integrated conversion over an infinite propagation from one point on the conformal boundary to another. For would-be local measurements, this is not a particularly strong conversion. 

The next temperature regime is the low temperature phase, but away from the critical temperature of the transition. We have not developed tools to focus on the actual transition, and its vicinity, but even the regular low temperature phase turns out to have a rich and interesting behaviour. Some of the speculations in the literature, related to this phase, concern the intriguing possibility of partial deconfinement \cite{Asplund:2008xd,Hanada:2016pwva,Berenstein:2018lrm} related to small black holes, also in this model \cite{Hanada:2019czd,Hanada:2020uvt}. Towards the transition point, the similarity of the vector model phase transition to a part of a phenomenological description of the matrix model/gauge theory phase diagram \cite{Alvarez-GaumeGomezLiuWadia2005} is also suggestive. The third order transition in this model was proposed to represent the correspondence point between a string gas and black holes \cite{Horowitz:1996nw,Chen:2021emg}.

Strong conversion in the low temperature phase is a definite and intriguing result of our analysis. One important contribution, eq. \eqref{eq:aligned}, increases as $T^3$ in the low temperature phase and we have argued against a complete cancellation of $T^3$ terms. Our approximations do not allow us to approach the actual phase transition, but we know the order of magnitude of mixing at $T=T_c/3 $, for example: $\mathcal{O}((T_c/3)^3)\sim \mathcal{O}(N)$. We believe that integrated conversion of this order, inverse to $G_N$, truly qualifies as strong conversion. Furthermore, this conversion occurs below the phase transition, in the low temperature phase. This lends support to the idea that the low temperature phase of the vector model represents a strong gravity source, be it a string-gas-like phase, a small black hole, or even a compact AdS star. 

Mixing correlators at extremely high temperature grow as $T^2$, see eq. \eqref{eq:extreme_high} just as the cold mixing correlators discussed above. This represents a return to a slower increase with temperature. Whatever the detailed explanation of this varying temperature dependence is, it must contain non-trivial physics. There is an intervening phase transition, which is likely to play a role, but a more detailed study would be needed to illuminate what is going on. 

Finally, we uncovered a simple result for the high temperature limit of the static mixed correlator, most easily described in terms of the momentum space correlator \eqref{eq:fdef} illustrated in figure~\ref{fig:plot}. Applying the standard relation between IR and UV boundary scales to the holographic bulk coordinate, we learn that there is a sharply defined central bulk region with strong conversion. The ground state scalar field and the temperature related temporal graviton dual to the boundary energy density are efficiently converted into one another in the central region.

\subsection{Conclusion}
In setting up this project, we were curious about the relation of mixing to other signals of strong gravity, especially evanescent modes. In vector models the presence of evanescent modes is related to non-uniformity of the eigenvalue density, a property that has not turned out to be equally important for mixing. The origins of these two effects is thus different, and we are led to regard them as complementary probes of strong gravity effects.

We have confirmed that mixed boundary correlators can probe details in the depth of the bulk, and illuminate what happens when the central bulk object is varied with changing temperature. It is clear that many more features can be probed. It will be especially interesting to probe the nature of the phase transition point more closely, but even the understanding of the low temperature phase can be improved substantially by developing techniques introduced in this paper further. Ultimately, with simple models of quantum black hole-like objects accessible to our study, it behoves us to shine a light on their properties.

\appendix
\section{Fundamental propagators} \label{app:fundamental}
The fundamental propagator can be read off from the action \eqref{eq:action} after some standard manipulations. First, we expand the fields in modes according to
\begin{equation}
\varphi^{a} = \sum_{n=-\infty}^{\infty}\sum_{l_{1},\ldots,l_{d-1}}v^{a}_{n,l_{1},\ldots,l_{d-1}}e^{i\omega_{n}\tau}Y_{l_{1},\ldots,l_{d-1}}(\phi_{1},\ldots,\phi_{d-1})\,,
\end{equation}
where $\omega_{n} = 2\pi n/\beta$ are Matsubara frequencies and $Y_{l_{1},\ldots,l_{d-1}}(\phi_{1},\ldots,\phi_{d-1})$ are hyperspherical harmonics. Then, we insert the identity matrix $\delta_{ab} = \sum_{i=1}^{N}\Psi^{a}_{i}(\Psi^{b}_{i})^{*}$ between the fields and the kinetic operator, where $\Psi^{a}_{i}$ is the eigenvector associated with the eigenvalue $\lambda_{i}$ of the matrix $\beta \alpha^{ab}$, yielding
\begin{equation}
S = \sum_{n,l_{1},\ldots,l_{d-1},i} (v^{a}_{n,l_{1},\ldots,l_{d-1}})^{*}\left((\omega_{n}+\lambda_{i}/\beta)^{2}+\frac{(l_{d-1}+(d-2)/2)^{2}}{R^{2}}\right)\Psi^{b}_{i}(\Psi^{a}_{i})^{*}v^{b}_{n,l_{1},\ldots,l_{d-1}}\,.
\end{equation}
The propagator $\Pi_{ab}(n,l)$ can now readily be read off:
\begin{equation}
\Pi_{ab}(n,l) = \sum_{i=1}^{N}\frac{\Psi^{a}_{i}(\Psi^{b}_{i})^{*}}{(\omega_{n}+\tfrac{\lambda_{i}}{\beta})^{2}+E_{l}^{2}}\,,
\end{equation}
where $E_{l} = l+(d-2)/2\equiv l+\sigma$ where we have set $R=1$. Transforming back to configuration space by expanding in Matsubara modes and hyperspherical Harmonics yields, after a rather technical computation,
\begin{equation}\label{eq:fundcorr}
\Pi_{ab}(\tau,\theta) = \sum_{i=1}^{N}\Psi^{a}_{i}(\Psi^{b}_{i})^{*}e^{-i\tau\frac{\lambda_{i}}{\beta}}\Bigg(\sum_{n=-\infty}^{\infty}\frac{e^{- i n \lambda_{i}}}{(\cosh(\tau+\beta n)-\cos\theta)^{\sigma}}\Bigg)\,,
\end{equation}
where we have used the underlying $SO(d-1)$ symmetry to express the spatial distance in terms of a single coordinate $\theta$. We have also removed a numerical prefactor so as to simplify expressions by effectively normalizing the fundamental fields.

\section{The gauge correction term to the mixing correlator}\label{app:correction}

Let us now compute the thermal gauge correction to the mixing correlator from the correction $T^{\mathrm{(th)}}_{00}$ to the energy density. Using the same trick as before, i.e., inserting the identity operator $\delta_{ab} = \sum_{i=1}^{N}(\Psi^{a}_{i})^{*}\Psi^{b}_{i}$ between the fields and the matrix $\alpha$. We thus obtain
\begin{align}
T^{\mathrm{(th)}}_{00}
&= \frac{5i}{3\beta} (\partial_{0}(\varphi^{\dagger})^{a}\sum_{i=1}^{N}\lambda_{i}(\Psi^{a}_{i})^{*}\Psi^{d}_{i}\varphi^{d} - (\varphi^{\dagger})^{a}\sum_{i=1}^{N}\lambda_{i}(\Psi^{a}_{i})^{*}\Psi^{d}_{i}\partial_{0}\varphi^{d})\notag\\
&+ \frac{1}{3\beta^{2}} (\varphi^{\dagger})^{a}\sum_{i=1}^{N}\lambda^{2}_{i}(\Psi^{a}_{i})^{*}\Psi^{d}_{i}\varphi^{d}\,,
\end{align}
where we used $\alpha (\Psi_{i})^{*} = \tfrac{\lambda_{i}}{\beta}(\Psi_{i})^{*}$ just like $\alpha \Psi_{i} = \tfrac{\lambda_{i}}{\beta}\Psi_{i}$ because of the Hermiticity of $\alpha$. The next step is to normal-order this expression, multiply it with $\normord{(\varphi^{\dagger})^{c}\varphi^{c}}$, and lastly take the expectation value. This produces a product of two Wick contractions in each term. We obtain
\begin{align}
&\frac{5i}{3\beta} (\partial_{0}\expval{(\varphi^{\dagger})^{a}\varphi^{c}}\sum_{i=1}^{N}\lambda_{i}(\Psi^{a}_{i})^{*}\Psi^{d}_{i}\expval{\varphi^{d}(\varphi^{\dagger})^{c}} - \expval{(\varphi^{\dagger})^{a}\varphi^{c}}\sum_{i=1}^{N}\lambda_{i}(\Psi^{a}_{i})^{*}\Psi^{d}_{i}\partial_{0}\expval{\varphi^{d}(\varphi^{\dagger})^{c}})\notag\\
&+ \frac{1}{3\beta^{2}} \expval{(\varphi^{\dagger})^{a}\varphi^{c}}\sum_{i=1}^{N}\lambda^{2}_{i}(\Psi^{a}_{i})^{*}\Psi^{d}_{i}\expval{\varphi^{d}(\varphi^{\dagger})^{c}}\,.
\end{align}
Now, making use of \eqref{eq:fundcorr} which we rewrite as
\begin{equation}
\sum_{i=1}^{N}\Psi^{a}_{i}(\Psi^{b}_{i})^{*}e^{-i\tau\frac{\lambda_{i}}{\beta}}\Bigg(\sum_{n=-\infty}^{\infty}\frac{e^{- i n \lambda_{i}}}{(\cosh(\tau+\beta n)-\cos\theta)^{\sigma}}\Bigg) \equiv \sum_{i=1}^{N}\Psi^{a}_{i}(\Psi^{b}_{i})^{*}f(\lambda_{i})\,,
\end{equation}
where
\begin{equation}
f(\lambda_{i}) = e^{-i\tau\frac{\lambda_{i}}{\beta}}\Bigg(\sum_{n=-\infty}^{\infty}\frac{e^{- i n \lambda_{i}}}{(\cosh(\tau+\beta n)-\cos\theta)^{\sigma}}\Bigg)\,,
\end{equation}
we obtain
\begin{align}
 \frac{5i}{3\beta} \left(\sum_{i=1}^{N}\lambda_{i}\partial_{0}f(\lambda_{i})f(-\lambda_{i}) - \sum_{i=1}^{N}\lambda_{i}f(\lambda_{i})\partial_{0}f(-\lambda_{i})\right)+ \frac{1}{3\beta^{2}} \sum_{i=1}^{N}\lambda^{2}_{i}f(\lambda_{i})f(-\lambda_{i})\,.
\end{align}
Now,
\begin{align}
\partial_{0}f(\lambda_{i}) &= \partial_{0}\left(e^{-i\tau\frac{\lambda_{i}}{\beta}}\Bigg[\sum_{n=-\infty}^{\infty}\frac{e^{- i n \lambda_{i}}}{(\cosh(\tau+\beta n)-\cos\theta)^{\sigma}}\Bigg]\right)\notag\\
&= -i\frac{\lambda_{i}}{\beta}f(\lambda_{i}) + e^{-i\tau\frac{\lambda_{i}}{\beta}}\Bigg[\sum_{n=-\infty}^{\infty}\frac{\sigma \sinh(\tau+\beta n)e^{- i n \lambda_{i}}}{(\cosh(\tau+\beta n)-\cos\theta)^{\sigma+1}}\Bigg]\,,
\end{align}
We thus obtain
\begin{align}
& \frac{10}{3\beta^{2}} \sum_{i=1}^{N}\lambda_{i}^{2}f(\lambda_{i})f(-\lambda_{i})\notag\\
&+\frac{10}{3\beta} \sum_{i=1}^{N}\lambda_{i}\Im\Bigg(f(\lambda_{i})e^{i\tau\frac{\lambda_{i}}{\beta}}\Bigg[\sum_{n=-\infty}^{\infty}\frac{\sigma \sinh(\tau+\beta n)e^{ i n \lambda_{i}}}{(\cosh(\tau+\beta n)-\cos\theta)^{\sigma+1}}\Bigg]\Bigg)\,.
\end{align}
In total, the contribution stemming from the gauge zero mode is thus
\begin{align}\label{eq:res}
&\frac{11}{3\beta^{2}} \sum_{i=1}^{N}\lambda_{i}^{2}g(\lambda_{i})g(-\lambda_{i})\notag\\
&+\frac{10}{3\beta} \sum_{i=1}^{N}\lambda_{i}\Im\Bigg(g(\lambda_{i})\Bigg[\sum_{n=-\infty}^{\infty}\frac{\sigma \sinh(\tau+\beta n)e^{ i n \lambda_{i}}}{(\cosh(\tau+\beta n)-\cos\theta)^{\sigma+1}}\Bigg]\Bigg)\,,
\end{align}
where
\begin{equation}
g(\lambda_{i}) = \sum_{m=-\infty}^{\infty}\frac{e^{- i m \lambda_{i}}}{(\cosh(\tau+\beta m)-\cos\theta)^{\sigma}}\,.
\end{equation}

Since
\begin{align}
&\Im\Bigg(g(\lambda_{i})\Bigg[\sum_{n=-\infty}^{\infty}\frac{\sigma \sinh(\tau+\beta n)e^{ i n \lambda_{i}}}{(\cosh(\tau+\beta n)-\cos\theta)^{\sigma+1}}\Bigg]\Bigg)\notag\\
&= \Im\Bigg(\Bigg[\sum_{n,m=-\infty}^{\infty}\frac{\sigma \sinh(\tau+\beta n)e^{ i (n-m) \lambda_{i}}}{(\cosh(\tau+\beta n)-\cos\theta)^{\sigma+1}(\cosh(\tau+\beta m)-\cos\theta)^{\sigma}}\Bigg]\Bigg)\notag\\
&=\Bigg[\sum_{n,m=-\infty}^{\infty}\frac{\sigma \sinh(\tau+\beta n)\sin((n-m) \lambda_{i})}{(\cosh(\tau+\beta n)-\cos\theta)^{\sigma+1}(\cosh(\tau+\beta m)-\cos\theta)^{\sigma}}\Bigg]\,,
\end{align}
and since $\lambda_{i}$ stems from an even distribution about 0 on $(-\pi,\pi)$, this contribution would vanish were it not for the prefactor $\lambda_{i}$ in \eqref{eq:res}. Hence, we keep all contributions:
\begin{align} \label{eq:th_uniform_density}
&\expval{T_{00}^{(th)}(\tau,x)\abs{\varphi}^{2}(0,0)} \notag\\
&\simeq \frac{11}{3\beta^{2}} \int\dd\lambda\sum_{n,m=-\infty}^{\infty}\frac{\rho(\lambda)\lambda^{2}e^{-i (n-m) \lambda}}{(\cosh(\tau+\beta n)-\cos\theta)^{\sigma}(\cosh(\tau+\beta m)-\cos\theta)^{\sigma}}\notag\\
&+\frac{10}{3\beta} \int\dd\lambda\Im\Bigg(\sum_{n,m=-\infty}^{\infty}\frac{\sigma \sinh(\tau+\beta n)\rho(\lambda)\lambda e^{ i (n-m) \lambda}}{(\cosh(\tau+\beta n)-\cos\theta)^{\sigma+1}(\cosh(\tau+\beta m)-\cos\theta)^{\sigma}}\Bigg)\,.
\end{align}

\acknowledgments

The work of B.S. is supported by the Swedish research council VR, contract DNR-2018-03803.

\bibliographystyle{JHEP}
\bibliography{Bibliography}

\providecommand{\href}[2]{#2}\begingroup\raggedright\begin{thebibliography}{10}

\bibitem{Grinberg:2020fdja}
M.~Grinberg and J.~Maldacena, \emph{{Proper time to the black hole singularity
  from thermal one-point functions}},
  \href{https://doi.org/10.1007/JHEP03(2021)131}{\emph{JHEP} {\bfseries 03}
  (2021) 131} [\href{https://arxiv.org/abs/2011.01004}{{\ttfamily
  2011.01004}}].

\bibitem{Rodriguez-Gomez:2021pfh}
D.~Rodriguez-Gomez and J.G.~Russo, \emph{{Correlation functions in finite
  temperature CFT and black hole singularities}},
  \href{https://arxiv.org/abs/2102.11891}{{\ttfamily 2102.11891}}.

\bibitem{Dodelson:2020lala}
M.~Dodelson and H.~Ooguri, \emph{{Singularities of thermal correlators at
  strong coupling}},
  \href{https://doi.org/10.1103/PhysRevD.103.066018}{\emph{Phys. Rev. D}
  {\bfseries 103} (2021) 066018}
  [\href{https://arxiv.org/abs/2010.09734}{{\ttfamily 2010.09734}}].

\bibitem{ReyRosenhaus2014}
S.-J.~Rey and V.~Rosenhaus, \emph{{Scanning Tunneling Macroscopy, Black Holes,
  and AdS/CFT Bulk Locality}},
  \href{https://doi.org/10.1007/JHEP07(2014)050}{\emph{JHEP} {\bfseries 07}
  (2014) 050} [\href{https://arxiv.org/abs/1403.3943}{{\ttfamily 1403.3943}}].

\bibitem{JevickiSuzuki2016}
A.~Jevicki and K.~Suzuki, \emph{{Thermofield Duality for Higher Spin Rindler
  Gravity}},  in \emph{JHEP\/} \cite{ShenkerYin2011}, 094,
  [\href{https://arxiv.org/abs/1508.07956}{{\ttfamily 1508.07956}}].

\bibitem{JevickiYoon2016}
A.~Jevicki and J.~Yoon, \emph{{Bulk from Bi-locals in Thermo Field CFT}},
  \href{https://doi.org/10.1007/JHEP02(2016)090}{\emph{JHEP} {\bfseries 02}
  (2016) 090} [\href{https://arxiv.org/abs/1503.08484}{{\ttfamily
  1503.08484}}].

\bibitem{AmadoSundborgThorlaciusWintergerst2017}
I.~Amado, B.~Sundborg, L.~Thorlacius and N.~Wintergerst, \emph{{Probing
  emergent geometry through phase transitions in free vector and matrix
  models}}, \href{https://doi.org/10.1007/JHEP02(2017)005}{\emph{JHEP}
  {\bfseries 02} (2017) 005}
  [\href{https://arxiv.org/abs/1612.03009}{{\ttfamily 1612.03009}}].

\bibitem{AmadoSundborgThorlaciusWintergerst2018}
I.~Amado, B.~Sundborg, L.~Thorlacius and N.~Wintergerst, \emph{{Black holes
  from large N singlet models}},
  \href{https://doi.org/10.1007/JHEP03(2018)075}{\emph{JHEP} {\bfseries 03}
  (2018) 075} [\href{https://arxiv.org/abs/1712.06963}{{\ttfamily
  1712.06963}}].

\bibitem{Hubeny:2006yu}
V.E.~Hubeny, H.~Liu and M.~Rangamani, \emph{{Bulk-cone singularities
  \textbackslash{}\& signatures of horizon formation in AdS/CFT}},
  \href{https://doi.org/10.1088/1126-6708/2007/01/009}{\emph{JHEP} {\bfseries
  01} (2007) 009} [\href{https://arxiv.org/abs/hep-th/0610041}{{\ttfamily
  hep-th/0610041}}].

\bibitem{TyukovWalkerWarner2018}
A.~Tyukov, R.~Walker and N.P.~Warner, \emph{{Tidal Stresses and Energy Gaps in
  Microstate Geometries}},
  \href{https://doi.org/10.1007/JHEP02(2018)122}{\emph{JHEP} {\bfseries 02}
  (2018) 122} [\href{https://arxiv.org/abs/1710.09006}{{\ttfamily
  1710.09006}}].

\bibitem{Bena:2018mpb}
I.~Bena, E.J.~Martinec, R.~Walker and N.P.~Warner, \emph{{Early Scrambling and
  Capped BTZ Geometries}},
  \href{https://doi.org/10.1007/JHEP04(2019)126}{\emph{JHEP} {\bfseries 04}
  (2019) 126} [\href{https://arxiv.org/abs/1812.05110}{{\ttfamily
  1812.05110}}].

\bibitem{Bena:2020iyw}
I.~Bena, A.~Houppe and N.P.~Warner, \emph{{Delaying the Inevitable: Tidal
  Disruption in Microstate Geometries}},
  \href{https://doi.org/10.1007/JHEP02(2021)103}{\emph{JHEP} {\bfseries 02}
  (2021) 103} [\href{https://arxiv.org/abs/2006.13939}{{\ttfamily
  2006.13939}}].

\bibitem{Martinec:2020cml}
E.J.~Martinec and N.P.~Warner, \emph{{The Harder They Fall, the Bigger They
  Become: Tidal Trapping of Strings by Microstate Geometries}},
  \href{https://doi.org/10.1007/JHEP04(2021)259}{\emph{JHEP} {\bfseries 04}
  (2021) 259} [\href{https://arxiv.org/abs/2009.07847}{{\ttfamily
  2009.07847}}].

\bibitem{Craps:2020ahu}
B.~Craps, M.~De~Clerck, P.~Hacker, K.~Nguyen and C.~Rabideau, \emph{{Slow
  scrambling in extremal BTZ and microstate geometries}},
  \href{https://doi.org/10.1007/JHEP03(2021)020}{\emph{JHEP} {\bfseries 03}
  (2021) 020} [\href{https://arxiv.org/abs/2009.08518}{{\ttfamily
  2009.08518}}].

\bibitem{Iliesiu:2018fao}
L.~Iliesiu, M.~Kolo\u{g}lu, R.~Mahajan, E.~Perlmutter and D.~Simmons-Duffin,
  \emph{{The Conformal Bootstrap at Finite Temperature}},
  \href{https://doi.org/10.1007/JHEP10(2018)070}{\emph{JHEP} {\bfseries 10}
  (2018) 070} [\href{https://arxiv.org/abs/1802.10266}{{\ttfamily
  1802.10266}}].

\bibitem{GobeilMaloneyNgWu2019}
Y.~Gobeil, A.~Maloney, G.S.~Ng and J.-q.~Wu, \emph{{Thermal Conformal Blocks}},
  \href{https://doi.org/10.21468/SciPostPhys.7.2.015}{\emph{SciPost Phys.}
  {\bfseries 7} (2019) 015} [\href{https://arxiv.org/abs/1802.10537}{{\ttfamily
  1802.10537}}].

\bibitem{Karlsson:2021duj}
R.~Karlsson, A.~Parnachev and P.~Tadi\'c, \emph{{Thermalization in Large-N
  CFTs}},  \href{https://arxiv.org/abs/2102.04953}{{\ttfamily 2102.04953}}.

\bibitem{Sundborg2000}
B.~Sundborg, \emph{{The Hagedorn Transition, Deconfinement and N=4 SYM
  Theory}}, \href{https://doi.org/10.1016/S0550-3213(00)00044-4}{\emph{Nucl.
  Phys.} {\bfseries B573} (2000) 349}
  [\href{https://arxiv.org/abs/hep-th/9908001}{{\ttfamily hep-th/9908001}}].

\bibitem{AharonyMarsanoMinwallaPapadodimasVan-Raamsdonk2004}
O.~Aharony, J.~Marsano, S.~Minwalla, K.~Papadodimas and M.~Van~Raamsdonk,
  \emph{{The Hagedorn - deconfinement phase transition in weakly coupled large
  N gauge theories}}, {\emph{Adv.Theor.Math.Phys.} {\bfseries 8} (2004) 603}
  [\href{https://arxiv.org/abs/hep-th/0310285}{{\ttfamily hep-th/0310285}}].

\bibitem{ShenkerYin2011}
S.H.~Shenker and X.~Yin, \emph{{Vector Models in the Singlet Sector at Finite
  Temperature}},  \href{https://arxiv.org/abs/1109.3519}{{\ttfamily
  1109.3519}}.

\bibitem{KlebanovPolyakov2002}
I.~Klebanov and A.~Polyakov, \emph{{AdS dual of the critical O(N) vector
  model}},
  \href{https://doi.org/10.1016/S0370-2693(02)02980-5}{\emph{Phys.Lett.}
  {\bfseries B550} (2002) 213}
  [\href{https://arxiv.org/abs/hep-th/0210114}{{\ttfamily hep-th/0210114}}].

\bibitem{FradkinVasiliev1987b}
E.~Fradkin and M.A.~Vasiliev, \emph{{On the Gravitational Interaction of
  Massless Higher Spin Fields}},
  \href{https://doi.org/10.1016/0370-2693(87)91275-5}{\emph{Phys.Lett.}
  {\bfseries B189} (1987) 89}.

\bibitem{FradkinVasiliev1987a}
E.~Fradkin and M.A.~Vasiliev, \emph{{Cubic Interaction in Extended Theories of
  Massless Higher Spin Fields}},
  \href{https://doi.org/10.1016/0550-3213(87)90469-X}{\emph{Nucl.Phys.}
  {\bfseries B291} (1987) 141}.

\bibitem{SezginSundell2002}
E.~Sezgin and P.~Sundell, \emph{{Massless higher spins and holography}},
  \href{https://doi.org/10.1016/S0550-3213(02)00739-3}{\emph{Nucl.Phys.}
  {\bfseries B644} (2002) 303}
  [\href{https://arxiv.org/abs/hep-th/0205131}{{\ttfamily hep-th/0205131}}].

\bibitem{Sundborg2001}
B.~Sundborg, \emph{{Stringy gravity, interacting tensionless strings and
  massless higher spins}},
  \href{https://doi.org/10.1016/S0920-5632(01)01545-6}{\emph{Nucl. Phys. Proc.
  Suppl.} {\bfseries 102} (2001) 113}
  [\href{https://arxiv.org/abs/hep-th/0103247}{{\ttfamily hep-th/0103247}}].

\bibitem{Haggi-ManiSundborg2000}
P.~Haggi-Mani and B.~Sundborg, \emph{{Free large N supersymmetric Yang-Mills
  theory as a string theory}}, {\emph{JHEP} {\bfseries 04} (2000) 031}
  [\href{https://arxiv.org/abs/hep-th/0002189}{{\ttfamily hep-th/0002189}}].

\bibitem{Giombi:2016ejx}
S.~Giombi, \emph{{Higher Spin CFT Duality}},  in \emph{{Proceedings,
  Theoretical Advanced Study Institute in Elementary Particle Physics: New
  Frontiers in Fields and Strings (TASI 2015): Boulder, CO, USA, June 1-26,
  2015}}, pp.~137--214, 2017,
  \href{https://doi.org/10.1142/9789813149441_0003}{DOI}
  [\href{https://arxiv.org/abs/1607.02967}{{\ttfamily 1607.02967}}].

\bibitem{BanerjeePapadodimRajuSamantrayShrivasta2020}
S.~Banerjee, K.~Papadodimas, S.~Raju, P.~Samantray and P.~Shrivastava, \emph{{A
  Bound on Thermal Relativistic Correlators at Large Spacelike Momenta}},
  \href{https://doi.org/10.21468/SciPostPhys.8.4.064}{\emph{SciPost Phys.}
  {\bfseries 8} (2020) 064} [\href{https://arxiv.org/abs/1902.07203}{{\ttfamily
  1902.07203}}].

\bibitem{Banerjee:2019iwd}
S.~Banerjee, J.~Engels\"oy, J.~Larana-Aragon, B.~Sundborg, L.~Thorlacius and
  N.~Wintergerst, \emph{{Quenched coupling, entangled equilibria, and
  correlated composite operators: a tale of two O(N) models}},
  \href{https://doi.org/10.1007/JHEP08(2019)139}{\emph{JHEP} {\bfseries 08}
  (2019) 139} [\href{https://arxiv.org/abs/1903.12242}{{\ttfamily
  1903.12242}}].

\bibitem{Hagedorn:1965st}
R.~Hagedorn, \emph{{Statistical thermodynamics of strong interactions at
  high-energies}}, {\emph{Nuovo Cim. Suppl.} {\bfseries 3} (1965) 147}.

\bibitem{Sundborg:1984uk}
B.~Sundborg, \emph{{Thermodynamics of Superstrings at High-energy Densities}},
  \href{https://doi.org/10.1016/0550-3213(85)90235-4}{\emph{Nucl. Phys. B}
  {\bfseries 254} (1985) 583}.

\bibitem{HarmarkWilhelm2018}
T.~Harmark and M.~Wilhelm, \emph{{Hagedorn Temperature of AdS$_5$/CFT$_4$ via
  Integrability}},
  \href{https://doi.org/10.1103/PhysRevLett.120.071605}{\emph{Phys. Rev. Lett.}
  {\bfseries 120} (2018) 071605}
  [\href{https://arxiv.org/abs/1706.03074}{{\ttfamily 1706.03074}}].

\bibitem{Harmark:2018red}
T.~Harmark and M.~Wilhelm, \emph{{The Hagedorn temperature of AdS$_5$/CFT$_4$
  at finite coupling via the Quantum Spectral Curve}},
  \href{https://doi.org/10.1016/j.physletb.2018.09.033}{\emph{Phys. Lett. B}
  {\bfseries 786} (2018) 53}
  [\href{https://arxiv.org/abs/1803.04416}{{\ttfamily 1803.04416}}].

\bibitem{Asplund:2008xd}
C.T.~Asplund and D.~Berenstein, \emph{{Small AdS black holes from SYM}},
  \href{https://doi.org/10.1016/j.physletb.2009.02.043}{\emph{Phys. Lett. B}
  {\bfseries 673} (2009) 264}
  [\href{https://arxiv.org/abs/0809.0712}{{\ttfamily 0809.0712}}].

\bibitem{Hanada:2016pwva}
M.~Hanada and J.~Maltz, \emph{{A proposal of the gauge theory description of
  the small Schwarzschild black hole in AdS$_5\times$S$^5$}},
  \href{https://doi.org/10.1007/JHEP02(2017)012}{\emph{JHEP} {\bfseries 02}
  (2017) 012} [\href{https://arxiv.org/abs/1608.03276}{{\ttfamily
  1608.03276}}].

\bibitem{Berenstein:2018lrm}
D.~Berenstein, \emph{{Submatrix deconfinement and small black holes in AdS}},
  \href{https://doi.org/10.1007/JHEP09(2018)054}{\emph{JHEP} {\bfseries 09}
  (2018) 054} [\href{https://arxiv.org/abs/1806.05729}{{\ttfamily
  1806.05729}}].

\bibitem{Hanada:2019czd}
M.~Hanada, A.~Jevicki, C.~Peng and N.~Wintergerst, \emph{{Anatomy of
  Deconfinement}}, \href{https://doi.org/10.1007/JHEP12(2019)167}{\emph{JHEP}
  {\bfseries 12} (2019) 167}
  [\href{https://arxiv.org/abs/1909.09118}{{\ttfamily 1909.09118}}].

\bibitem{Hanada:2020uvt}
M.~Hanada, H.~Shimada and N.~Wintergerst, \emph{{Color Confinement and
  Bose-Einstein Condensation}},
  \href{https://arxiv.org/abs/2001.10459}{{\ttfamily 2001.10459}}.

\bibitem{Alvarez-GaumeGomezLiuWadia2005}
L.~Alvarez-Gaume, C.~Gomez, H.~Liu and S.~Wadia, \emph{{Finite temperature
  effective action, AdS(5) black holes, and 1/N expansion}},
  \href{https://doi.org/10.1103/PhysRevD.71.124023}{\emph{Phys. Rev. D}
  {\bfseries 71} (2005) 124023}
  [\href{https://arxiv.org/abs/hep-th/0502227}{{\ttfamily hep-th/0502227}}].

\bibitem{Horowitz:1996nw}
G.T.~Horowitz and J.~Polchinski, \emph{{A Correspondence principle for black
  holes and strings}},
  \href{https://doi.org/10.1103/PhysRevD.55.6189}{\emph{Phys. Rev. D}
  {\bfseries 55} (1997) 6189}
  [\href{https://arxiv.org/abs/hep-th/9612146}{{\ttfamily hep-th/9612146}}].

\bibitem{Chen:2021emg}
Y.~Chen and J.~Maldacena, \emph{{String scale black holes at large $D$}},
  \href{https://arxiv.org/abs/2106.02169}{{\ttfamily 2106.02169}}.

\end{thebibliography}\endgroup
\end{document}